\documentclass[aps,prb,twocolumn,superscriptaddress]{revtex4-1}

\usepackage{graphicx}
\usepackage{graphics}
\usepackage{subfigure}
\usepackage{amssymb,amsmath}
\usepackage{epsfig}
\usepackage{color}

\begin{document}


\title{From the artificial atom to the Kondo-Anderson model: orientation dependent magneto-photoluminescence of charged excitons in InAs quantum dots}


\author{B. Van Hattem}
\email[]{bv252@cam.ac.uk} \affiliation{Cavendish Laboratory, University of Cambridge, J. J. Thomson Avenue, CB3 0HE Cambridge, United Kingdom}
\author{P. Corfdir}
\affiliation{Cavendish Laboratory, University of Cambridge, J. J. Thomson Avenue, CB3 0HE Cambridge, United Kingdom}
\author{P. Brereton}
\altaffiliation{Present address: Navy Research Laboratory, Electronic Materials Branch,Code 6877 4555 Overlook Ave. SW, Washington, DC 20375} \affiliation{Cavendish Laboratory, University of Cambridge, J. J. Thomson Avenue, CB3 0HE Cambridge, United Kingdom}
\author{P. Pearce}
\affiliation{Cavendish Laboratory, University of Cambridge, J. J. Thomson Avenue, CB3 0HE Cambridge, United Kingdom}
\author{A. M. Graham}
\affiliation{Cavendish Laboratory, University of Cambridge, J. J. Thomson Avenue, CB3 0HE Cambridge, United Kingdom}
\author{M. J. Stanley}
\affiliation{Cavendish Laboratory, University of Cambridge, J. J. Thomson Avenue, CB3 0HE Cambridge, United Kingdom}
\author{M. Hugues}
\altaffiliation{Present address: Centre de Recherche sur l'H\'et\'ero-Epitaxie et ses Applications, Centre National de la Recherche Scientifique (CRHEA-CNRS), Parc Sophia Antipolis, Rue Bernard Gr\'egory, 06560 Valbonne, France} \affiliation{Department of Electronic and Electrical Engineering, University of Sheffield, Mappin Street, Sheffield S1 3JD, United Kingdom}
\author{M. Hopkinson}
\affiliation{Department of Electronic and Electrical Engineering, University of Sheffield, Mappin Street, Sheffield S1 3JD, United Kingdom}
\author{R. T. Phillips.}
\affiliation{Cavendish Laboratory, University of Cambridge, J. J. Thomson Avenue, CB3 0HE Cambridge, United Kingdom}


\date{\today}

\begin{abstract}
We present a magneto-photoluminescence study on neutral and charged excitons confined to InAs/GaAs quantum dots. Our investigation relies on a confocal microscope that allows arbitrary tuning of the angle between the applied magnetic field and the sample growth axis. First, from experiments on neutral excitons and trions, we extract the in-plane and on-axis components of the Land\'e tensor for electrons and holes in the $s$-shell. Then, based on the doubly negatively charged exciton magneto-photoluminescence we show that the $p$-electron wave function spreads significantly into the GaAs barriers. We also demonstrate that the $p$-electron $g$-factor depends on the presence of a hole in the $s$-shell. The magnetic field dependence of triply negatively charged excitons photoluminescence exhibits several anticrossings, as a result of coupling between the quantum dot electronic states and the wetting layer. Finally, we discuss how the system evolves from a Kondo-Anderson exciton description to the artificial atom model when the orientation of the magnetic field goes from Faraday to Voigt geometry.

\end{abstract}
\graphicspath{{figures/}}

\maketitle

\section{Introduction}

In the last fifteen years, quantum dots (QDs) have attracted a lot of interest from the scientific community as they allow for the development of a new class of quantum optoelectronic devices including single photon emitters\cite{Michler2000}, photon entangled pairs emitters\cite{Akopian2006,Stevenson2006} or all-optical logic gates.\cite{Li2003} Furthermore, recognition that the behaviour of QDs departs from the ``artificial atom'' model opens new areas of investigation involving the coupling between the discrete electronic states of the QD with a continuum.\cite{Sasaki2000,Karrai2004,Dalgarno2008,Kleemans2010} One of the most versatile ways of investigating the fine structure and the spin properties of electronic excitations confined to QDs is to study their response to a magnetic field. \cite{Bayer2002,Steffan2002,Cade2006,Sanada2009,Abbarchi2010,Witek2011} Magneto-photoluminescence experiments on excitons and trions have also proved successful in aiding estimation of the shape,\cite{Schulhauser2002,Fu2010} the alloy composition,\cite{Sheng2008a} and the symmetry of QDs.\cite{Sallen2011,Oberli2012} Less is known about the magneto-photoluminescence of more negatively charged excitonic complexes.\cite{Karrai2004,Kazimierczuk2011} Still, the spectroscopy of such highly charged excitons makes it possible to study the interaction between charge carriers occupying different states in the dot.\cite{Ediger2007b,Ediger2007a,Kazimierczuk2011} In addition, when three or more electrons are confined in the QD, applying a magnetic field may hybridize the QD electronic states with the Landau levels of the two-dimensional wetting layer.\cite{Karrai2004}

In this work, we present a magneto-photoluminescence (PL) study of charged excitons confined in InAs$/$GaAs QDs. Our experiment permits arbitrary tuning of the angle between the applied magnetic field and the sample growth axis. This allows the study of the evolution of the diamagnetic shift of the excitonic complex with respect to the orientation of the field, providing information on the electron-hole correlation length of the exciton wave function, and the extraction of the in-plane and on-axis components of the electron and hole Land\'e $g$-factors. In the peculiar case of the doubly negatively charged exciton $X^{2-}$, we deduce from its magneto-PL that the $p$-shell electron wave function spreads significantly inside the GaAs barriers and that this spreading strongly depends on the presence of a hole in the $s$-shell. The triply negatively charged exciton $X^{3-}$ is found to show PL that strongly deviates from the magnetic field dependence seen for the neutral exciton, the trions and the $X^{2-}$. As in the work of Karrai {\itshape et al.}, \cite{Karrai2004} we attribute this observation to the hybridization between the QDs electronic states with the wetting layer and we describe the evolution of this coupling as the magnetic field is rotated.

The paper is organized as follows. In Section II, we present the structure of our QD sample and we describe our confocal magneto-PL apparatus. In Section III, we identify by PL the different charge states of the exciton. In Section IV, we discuss the magneto-PL of the neutral exciton, the trions and the $X^{2-}$. Based on the magnetic field dependence of the emission lines, we deduce in particular in Sections IV.A and IV.B the shape of the exciton wave function and the electron and hole Land\'e tensors, respectively. In Section V, we present the magnetic field dependence of the $X^{3-}$ and we describe how the coupling of the final state of the recombination with the wetting layer affects the exciton recombination energies. We finally present our conclusions in Section VI.

\section{Experimental Details}

\begin{figure}[htbp]
\includegraphics[scale=0.4]{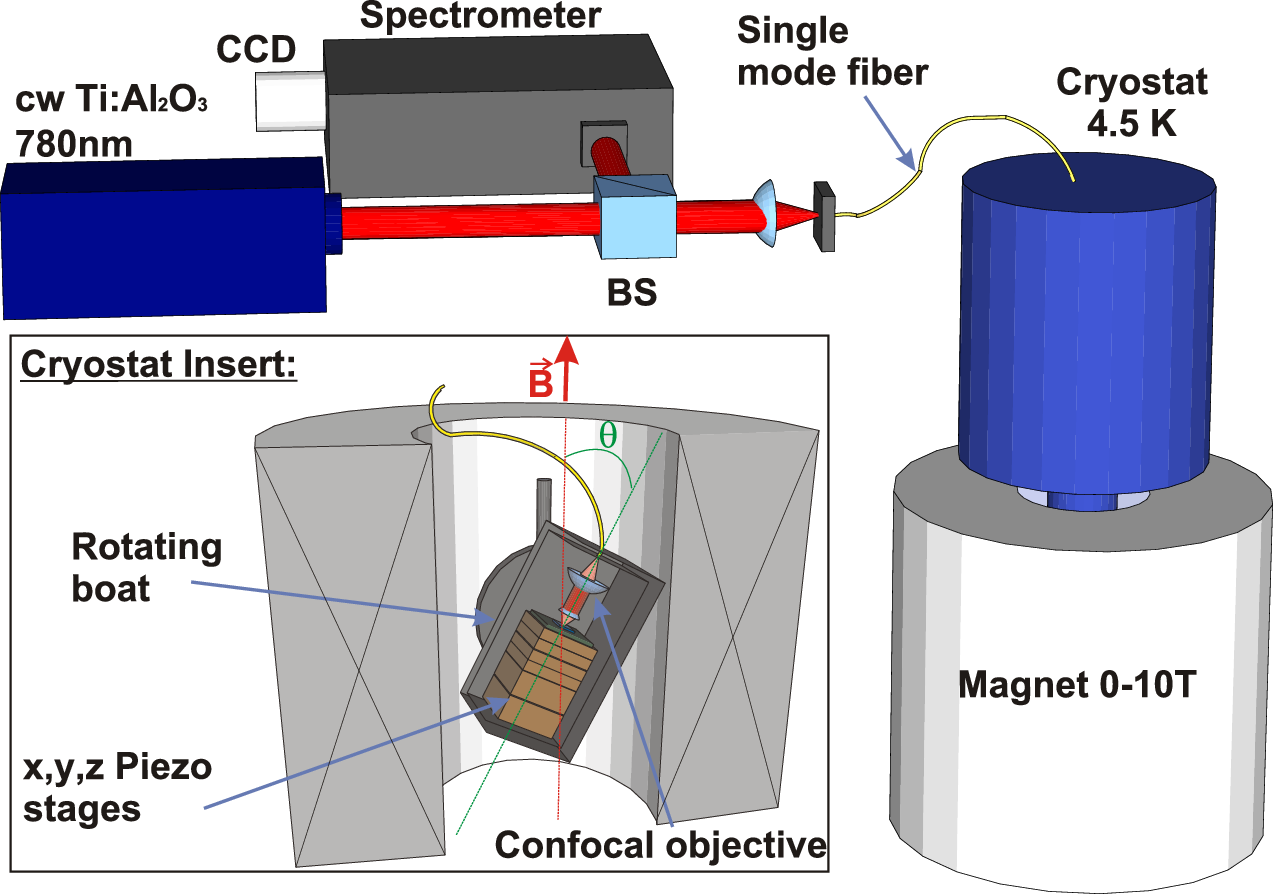}
\caption {(colour online) Confocal magneto-photoluminescence set-up. The applied magnetic field is always along the vertical direction. Inset: the sample and the microscope objective are mounted on a rotatable support. The angle $\theta$ can take any value between 0 and $90^o$}. \label{fig:Figure1}
\end{figure}

The InAs/GaAs QD sample has been grown by molecular beam epitaxy. The QDs are obtained by the Stranski-Krastanov growth mode transition and are embedded in a field effect structure. The QD layer is located 25 nm above a $n$-doped GaAs substrate and 205 nm below a Schottky contact layer. We deposited on top of the structure a 40 nm gold mask patterned with apertures with 400 nm diameter. In order to access the different charge states of the QD, a voltage ranging between -0.8 and 0.15 V was applied to the Schottky structure. For all PL experiments, the sample was excited with a cw Ti:Al$_{2}$O$_{3}$ laser operating at 780 nm. In the case of PL experiments at zero magnetic field, the emitted light was analysed by a triple spectrometer operating in additive mode (spectral resolution better than 20 $\mu$eV) and was detected by a Peltier-cooled CCD. Magneto-PL experiments were carried out at 4.5 K in a confocal set-up, in which the confocal head is rotatable with respect to the field, as showed in Figure \ref{fig:Figure1}.\cite{Kehoe2010} This allows setting the angle $\theta$ between the sample growth axis and the applied magnetic field to any arbitrary value between 0 (Faraday geometry) and 90$^o$ (Voigt geometry). Magnetic fields up to B = 10 T have been obtained with a `dry' superconducting magnet. In the present study, we only consider $B = (B_x = B\sin(\theta),B_y=0,B_z = B\cos(\theta))$, where the $x$, $y$ and $z$ directions correspond to the [1$\overline{1}$0], [110] and [001] crystallographic axes of the QD sample, respectively. For PL experiments carried out in magnetic field the light was analysed by a single monochromator with spectral resolution better than 150 $\mu$eV. While the conclusions made in this paper are based on the magneto-PL characterization of more than 25 QDs from the same sample, we have chosen, for clarity reasons, only to  present here the results obtained on a single representative QD.

\section{Identification of the charge states of the exciton}

\begin{figure}[h]
\includegraphics[scale=0.55]{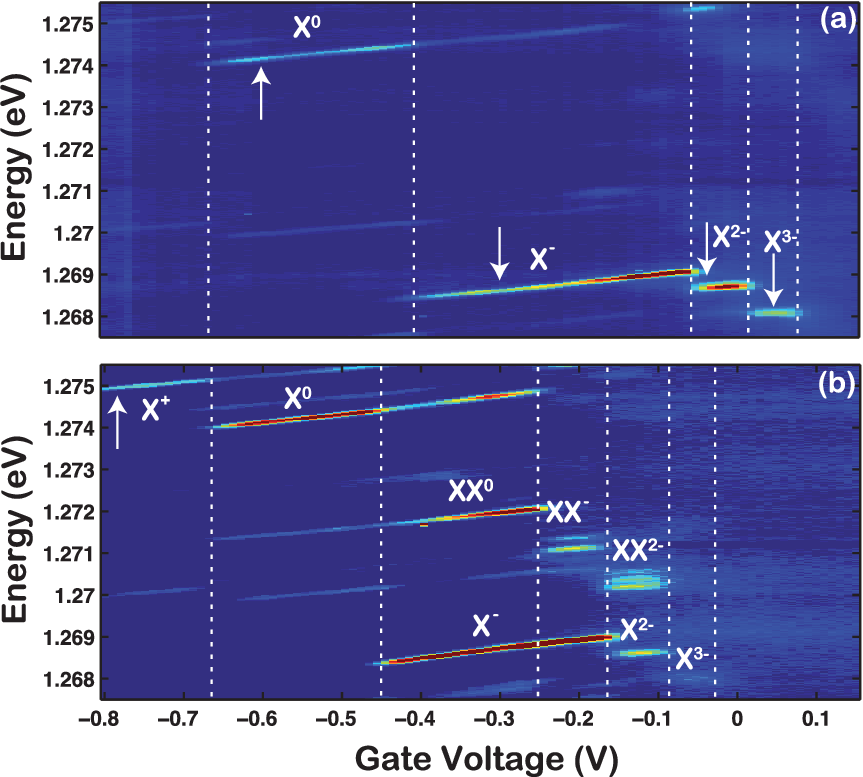}
\caption {(colour online): QD emission spectra as a function of the gate voltage, for an excitation density of $5\times10^9\mu$W/cm${^2}$ (a) and $60\times10^9 \mu$W/cm${^2}$ (b). Vertical dotted lines indicate the different charging steps. Arrows show at which voltage the magneto-PL scans in Figures \ref{fig:Figure5}, \ref{fig:Figure6} and \ref{fig:Figure7} have been taken.} \label{fig:Figure2}
\end{figure}

We show in Figure \ref{fig:Figure2} the QD emission spectra at zero magnetic field for applied gate voltages between -0.8 to 0.15 V. We attribute the abrupt changes in the QD PL energy to the charging of the QD.\cite{Warburton2000} Using the method of Ediger {\itshape et al.} \cite{Ediger2007a}, we identify  the emission from the neutral exciton $X^{0}$ and from the singly, doubly and triply negatively charged excitons $X^{-}$, $X^{2-}$ and $X^{3-}$, respectively. Applying gate voltages larger than 0.1 V results in the filling of wetting layer states, making it impossible to observe the emission from QD exciton states more negatively charged than $X^{3-}$. \cite{Warburton2000,Findeis2001,Finley2001,Poem2007} When increasing the excitation density, additional lines become visible [Figure \ref{fig:Figure2}(b)]. We attribute them to the recombination of neutral and charged biexciton complexes\cite{Poem2007} and to the emission from the positively charged trion $X^{+}$.\cite{Ediger2005} Finally, we emphasize that the gate voltages at the charging steps are excitation dependent, which we ascribe to the modification of the effective electrostatic potential felt by the QD in presence of excess charges.

\section{Magneto-Photoluminescence of the $X^{+}$, $X^{0}$, $X^{-}$ and $X^{2-}$ Quantum Dot States}

In the presence of a magnetic field, the emission lines from the $X^{+}$ to $X^{2-}$ complexes split and shift. The QD emission energies $E$ follow $E(B) = E(0) + \gamma_{1}B+\gamma_{2}B^{2}$.\cite{Walck1998} In the case of $X^0$, $\gamma_{1}B$  describes the Zeeman effect and $\gamma_{2}$ is the diamagnetic coefficient. We  first discuss in Section III.A the evolution of $\gamma_{2}$ with respect to $\theta$ and to the charge-state of the exciton. Then, we extract in Section III.B the electron and hole $g-$factors and the exchange energies for the different complexes.

\subsection{Diamagnetic shift of the neutral and charged exciton complexes}
\label{sec:diamagnetic}

\begin{figure}[htbp]
\includegraphics[scale=0.33]{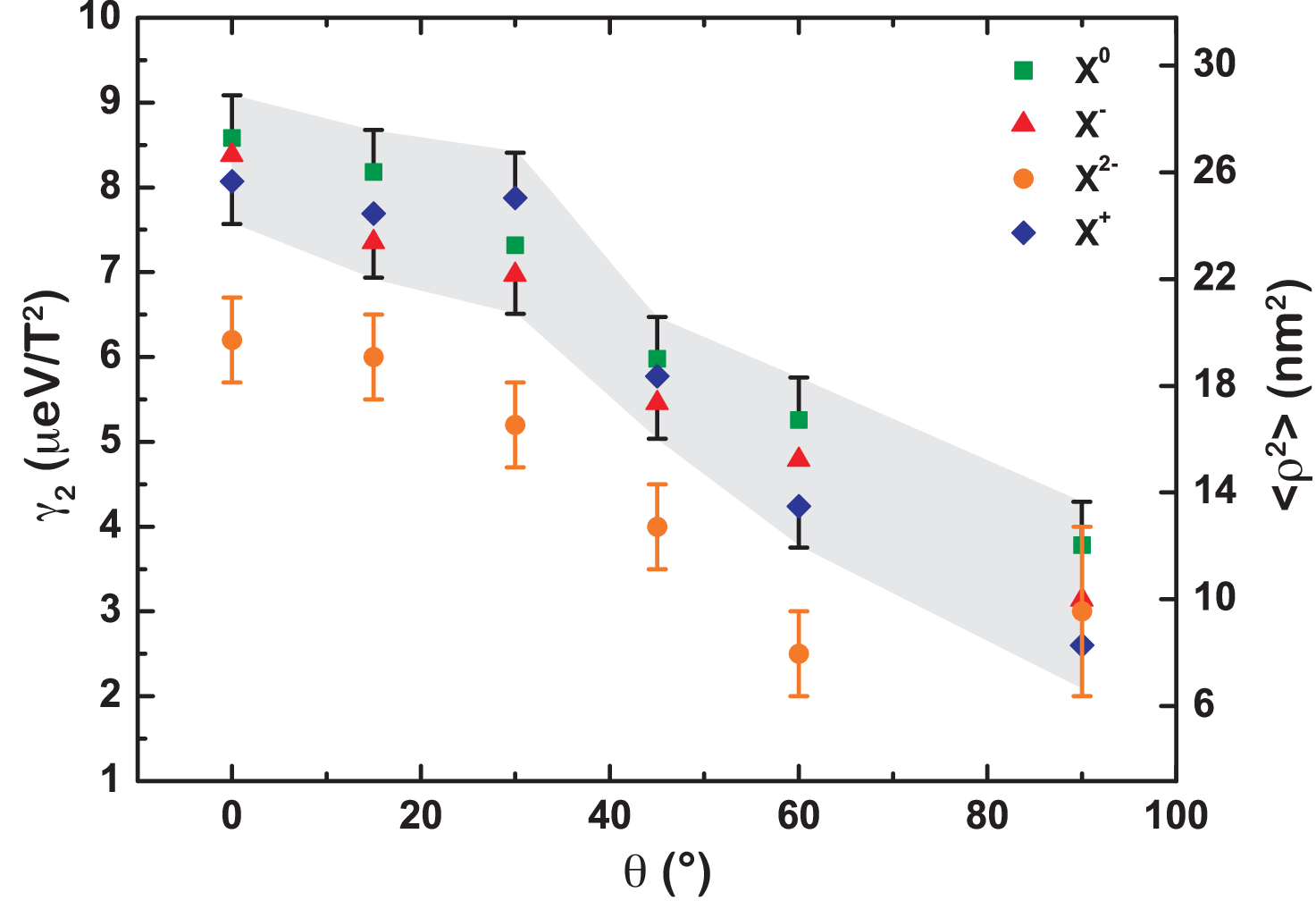}
\caption {(color online): Diamagnetic shift $\gamma_2$ as a function of $\theta$ for an excitation density of $5\times10^9\mu$W/cm${^2}$ (for $X^{+}$ an excitation density of $60\times10^9 \mu$W/cm${^2}$ was required). Experiments on $X^{+}$, $X^{0}$, $X^{-}$, $X^{2-}$ have been taken with gate voltages of -0.78, -0.60,-0.30 and -0.04 V, respectively. On the right scale, the corresponding values for the square of the $X^{0}$ coherence length $\langle\rho^2\rangle$ are shown. The grey shaded area shows the uncertainty on the $\gamma_2$ values extracted for $X^0$, $X^+$ and $X^-$.} \label{fig:Figure3}
\end{figure}

We show in Figure \ref{fig:Figure3} the angle dependence of $\gamma_{2}$ for $X^{0}$, $X^{+}$, $X^{-}$ and $X^{2-}$.  The diamagnetic shift of $X^0$ is given by $\gamma_{2}^{X^0} = e^2 {\langle\rho^2\rangle} / 8\mu$,\cite{Nash1989,Walck1998} where $\mu$ and $\sqrt{\langle\rho^2\rangle}$ are the exciton reduced mass and the exciton correlation length perpendicular to the axis of the applied magnetic field, respectively. As shown in Figure \ref{fig:Figure3}, $\gamma_{2}^{X^0}$ decreases from 8.5 to 4.0 $\mu$eV/T$^{2}$ when going from Faraday to Voigt geometry. Even if the hole mass anisotropy contributes slightly to the decrease of $\gamma_{2}^{X^0}$ when increasing $\theta$, the $\theta$-dependence of $\gamma_{2}^{X^0}$ arises mainly from the geometry of the investigated QD. The corresponding $\theta$-dependence of $\sqrt{\langle\rho^2\rangle}$ indicates that this QD exhibits a lens-shaped form. Clearly the shape of the exciton wave function is imposed by the QD geometry, which confirms that the investigated excitons are strongly confined in the QD. For charged complexes, $\gamma_{2}$ is given by the diamagnetic shift of both the initial and the final states of the transition: it consequently provides some information on the relative spatial extent of these two states.\cite{Schulhauser2002,Fu2010} For self-assembled InAs/GaAs QDs, $\gamma_{2}$ for $X^-$ may be equal to or smaller than  $\gamma_{2}^{X^0}$, depending on the size of the QD.\cite{Schulhauser2002} As already detailed by Fu {\itshape et al.} \cite{Fu2010}, for the smallest QDs, the wavefunction of the $s$-electron spreads significantly in the barriers. The presence of a hole in the QD reduces the electron delocalization, as a consequence of Coulomb attraction. Due to the increase in spatial extent for the $s$-electron wavefunction after recombination of the $X^-$, the corresponding emission line shows a reduced $\gamma_{2}$ compared to  $\gamma_{2}^{X^0}$. Although the previous discussion in principle also applies to the case of $X^+$, its  $\gamma_{2}$ should not be much affected by the presence of an electron in the QD, since the large effective mass of the heavy hole prevents it from spreading into the barrier. As shown in Figure 3, we measure in Faraday geometry similar  $\gamma_{2}$ for $X^0$, $X^+$ and $X^-$. This indicates that both the $s$-electron and the s-hole are strongly localized in the QD. In contrast to this, for $X^{2-}$ we observe a much smaller $\gamma_{2}$. This shows that the $p$-electron wave function is sensitive to the presence of a hole in the QD: in the initial state of the transition, the hole binds the $p$-electron to the QD; after recombination the $p$-electron is more weakly confined and its wave function spreads more into the GaAs barriers. Finally, compared to $X^{0}$ and the trions, $\gamma_{2}$ for the $X^{2-}$ shows a reduced anisotropy that arises from the delocalization of the $p$-electron inside the GaAs barriers.

\begin{figure}[htbp]
\includegraphics[scale=0.33]{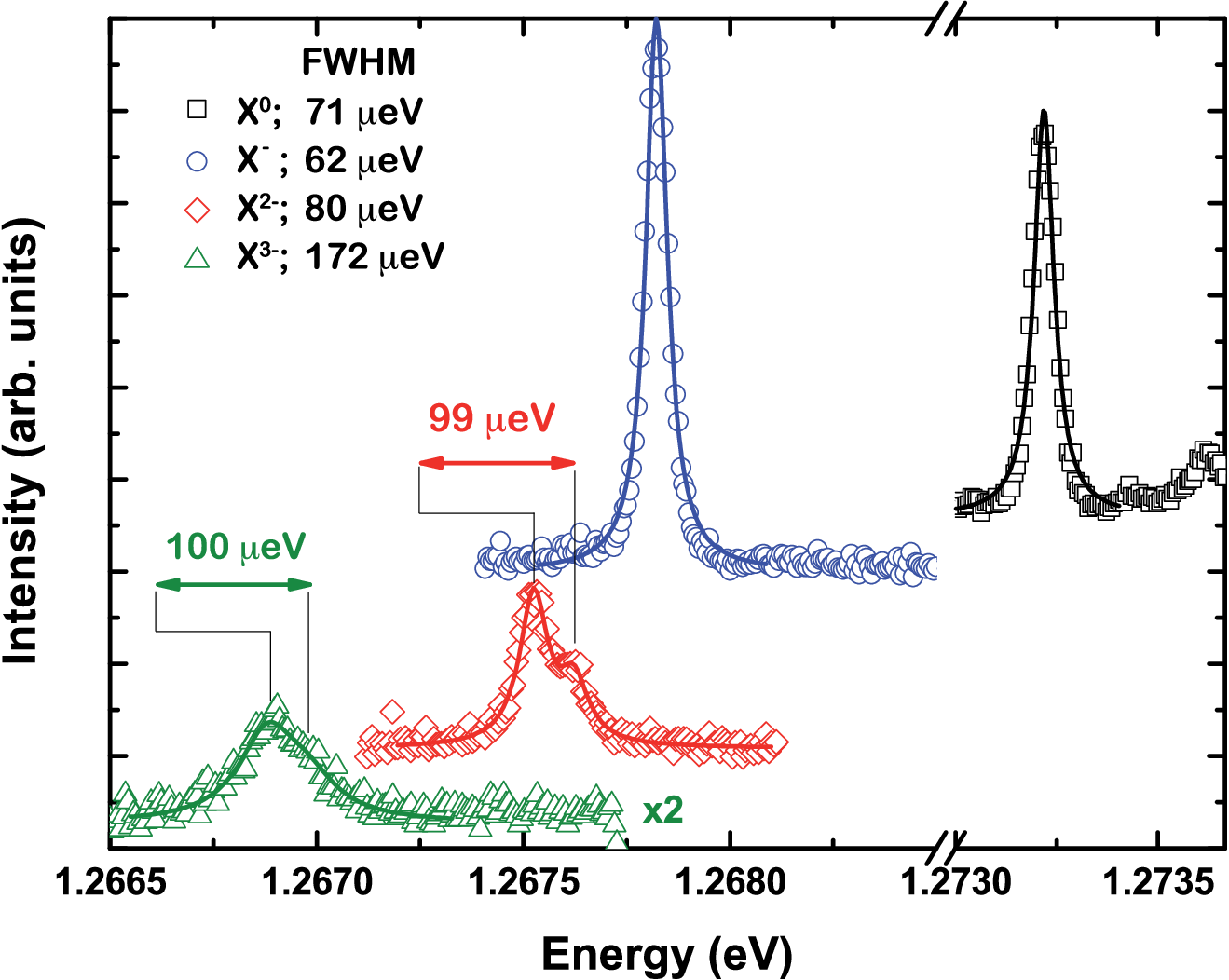}
\caption {(colour online): PL spectra of the $X^0$, $X^{-}$, $X^{2-}$ and $X^{3-}$ complexes (squares, circles, diamonds and triangles, respectively). Solid lines show the result of fitting with Lorentzian curves.} \label{fig:Figure4}
\end{figure}

\subsection{Magneto-photoluminescence of neutral and singly charged excitons}
\label{sec:neutral_trion}

\begin{figure}[htbp]
\includegraphics[scale=0.33]{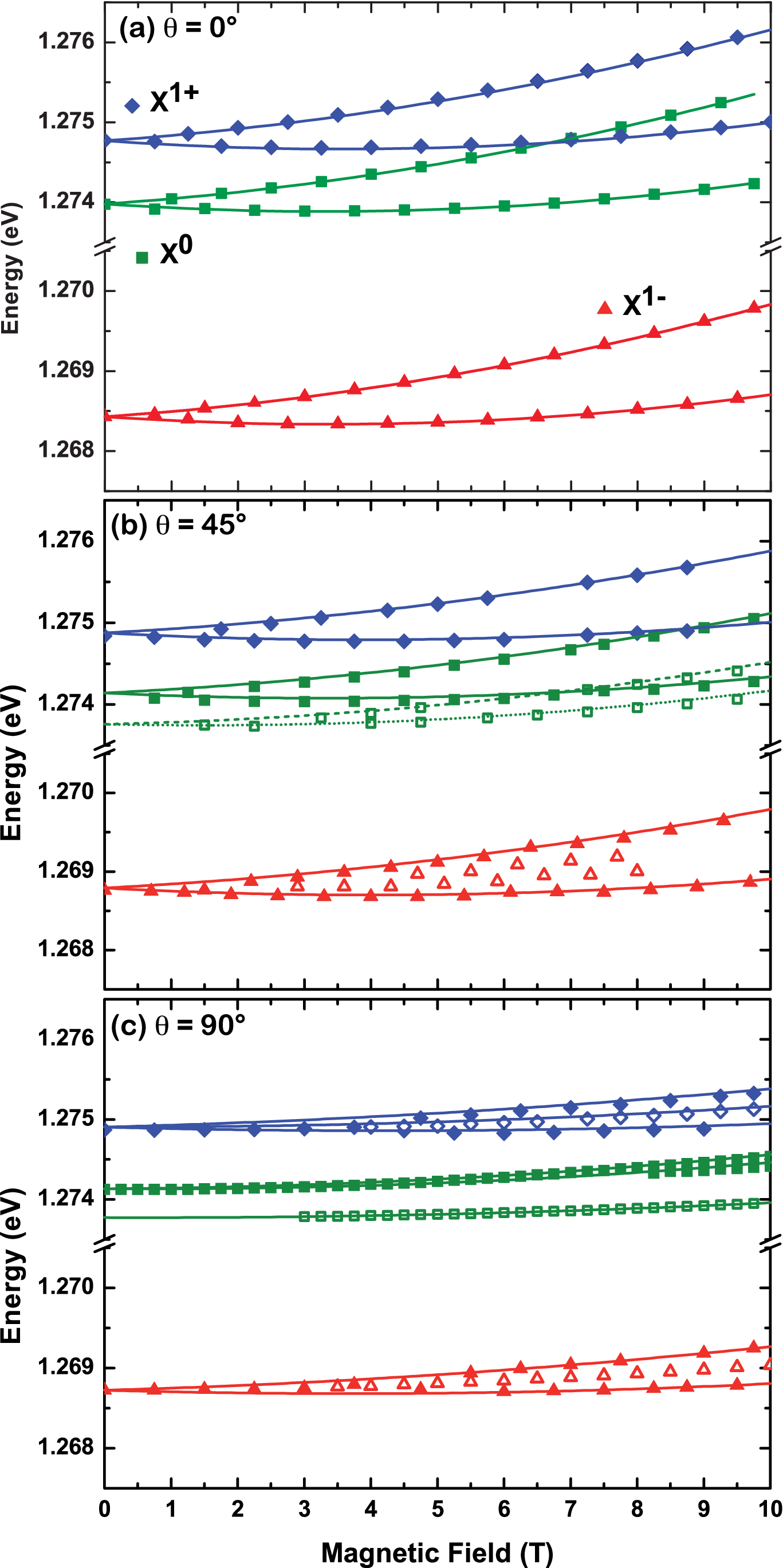}
\caption {(colour online): Magnetic field-dependence of the PL energies of the $X^{+},X^{0}$ and $X^{-}$  complexes (diamonds, squares and triangles respectively) at 4.5 K for $\theta=0^o, 45^o$ and $90^o$. Full and open symbols correspond to the emission from the primarily bright and dark states, respectively. In Faraday geometry, the $X^{+}$, $X^{0}$ and $X^{-}$   magneto-PL spectra have been recorded at gate voltages of -0.78, -0.60 and -0.30V respectively (comparable voltages were used for $\theta\neq 0$). Solid and dashed lines show the result of the fitting procedure of the emission energies of the predominantly bright and dark states, respectively.} \label{fig:Figure5}
\end{figure}

We display in Figure \ref{fig:Figure4} emission spectra from the neutral and negatively charged exciton complexes at zero magnetic field. The $X^0$ full width at half maximum (FWHM) is 71 $\mu$eV and no splitting is resolved. We therefore estimate that the anisotropic exchange splitting $|c_x+c_y|/2$ between the two bright states of $X^0$ is not larger than 30 $\mu$eV ($c_x$ and $c_y$ characterize the in-plane electron-hole spin-spin interaction along the [1$\overline{1}$0] and [110] crystal axes, respectively),\cite{Steffan2002,Phillips2003} which indicates that the symmetry of the QD is nearly $D_{2d}$. Coming to the anisotropic exchange splitting $|c_x-c_y|/2$ between the two dark $X_0$ states, it is of the order of $\mu$eV\cite{Poem2010} we therefore neglect it, as is usually done when discussing the magneto-PL of InAs QDs.\cite{Bayer2002} No fine structure is expected for the trion emission because of electron spin pairing and, for a given excitation density, the $X^-$ emission is narrower than the $X^0$. Note that the 62 $\mu$eV linewidth measured for the $X^{-}$ emission is likely to be given by spectral diffusion effects.\cite{Favero2007} The magnetic field-dependences of $X^{+}$, $X^{0}$ and $X^{-}$ emission energies for $\theta$ equal to 0, 45 and 90$^{o}$ are shown in Figure \ref{fig:Figure5}. In Faraday geometry, the emission from the $X^{0}$ bright states splits into two lines, whose energies $E_{F}^{X^0}$ follow:\cite{Bayer2002,Steffan2002}

 \begin{equation}
\label{eqn:faraday_neutral} E_{F}^{X^0}(B)=E_0+\frac{c_z}{4}\pm\frac{1}{2}\mu_{B}B|g_e^{sz}+g_h^{z}|+\gamma_{2}^{X^0}(0^o)B^2.
\end{equation}

 $g_{e}^{sz}$ and $g_{h}^{z}$ are the $s$-electron and hole Land\'e factors along the dot axis, respectively, and $c_z/2$ is the isotropic exchange splitting.\cite{Steffan2002,Phillips2003} Away from Faraday geometry, the in-plane component of the magnetic field introduces some mixing between the $X^0$ bright and dark states. Accordingly, two additional emission lines are observed for fields larger than about 2 T [Figure \ref{fig:Figure5}(b,c)]. In the specific case of Voigt geometry, the eigenenergies $E_{V,B}^{X^0}$ and  $E_{V,D}^{X^0}$ of the primarily bright and dark states, respectively, are given by:\cite{Bayer2002,Toft2007}

\begin{equation}
\label{eqn:voigt_b_neutral}
\begin{split}
E_{V,B}^{X^0}(B)&=E_0+ \frac{c_x}{4} \pm\frac{1}{2}\sqrt{\frac{(c_z- c_y)^2}{4}+ \mu_{B}^2B^2(g_h^{x}+g_e^{sx})^2}\\
& +\gamma_{2}^{X^0}(90^o)B^2,
\end{split}
\end{equation}

and

\begin{equation}
\label{eqn:voigt_d_neutral}
\begin{split}
E_{V,D}^{X^0}(B)&=E_0- \frac{c_x}{4}\pm\frac{1}{2}\sqrt{\frac{(c_z+ c_y)^2}{4}+\mu_{B}^2B^2(g_h^{x}- g_e^{sx})^2}\\
&+\gamma_{2}^{X^0}(90^o)B^2,
\end{split}
\end{equation}

 where $g_{e}^{sx}$ and $g_{h}^{x}$ are the $s$-electron and hole $g$-factors along [1$\overline{1}$0], respectively. While the data taken in Faraday and Voigt geometries were fitted using Equations (1-3), the data taken at intermediate $\theta$ were analyzed solving numerically the van Kesteren Hamiltonian.\cite{VanKesteren1990,Steffan2002} Using the magnetic field dependence of $X^0$ emission energies for $\theta$ equal to 30, 45, 60 and 90$^o$, we extract $c_z=718 \pm54$ $\mu$eV. This rather large $c_z$ is in agreement with the fact that the investigated $X^0$ is strongly confined, $c_z$ having been shown to scale like the inverse of the QD volume.\cite{Bayer2002} Coming to the trions, their emission energies in Faraday geometry are given by:\cite{Tischler2002}

\begin{figure*}[h!t]
  \includegraphics[width=\textwidth]{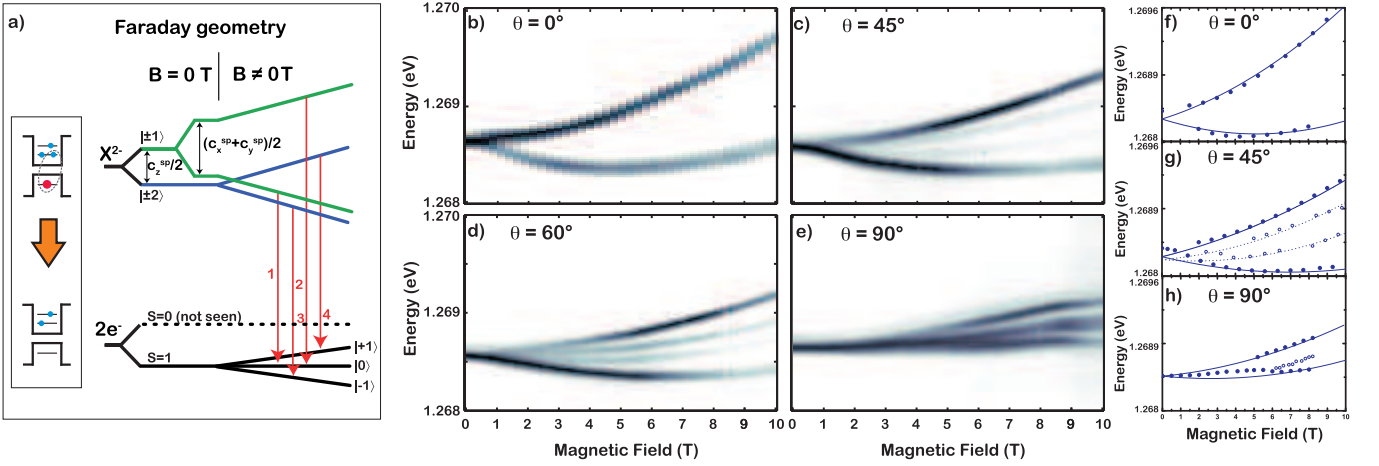}

\caption {(colour online): (a): Schematic representation of the $X^{2-}$ fine structure. The red vertical arrows show the four optically allowed transitions in Faraday geometry. (b-e) Magneto-PL of the $X^{2-}$ for  $\theta=0^o$ (b), $45^o$ (c), $60^o$ (d) and $90^o$ (e). (b) In Faraday geometry, two emission lines are resolved. (c-e) For non-zero $\theta$, the mixing between bright and dark states makes it possible to observe four distinct emission lines. In this tilted state, for fields larger than about 8 T the asymmetry of diamagnetic forces in the apparatus is sufficient to alter slightly the optical alignment of the system, resulting in a decrease in emission intensity for all emission lines. In (e) the ``jump" occurring at 8 T is caused by a magnetic field-induced change in the electrostatic environment of the QD (f-h): Experimental data points (blue circles) and parabolic fitting (blue lines) corresponding to the data shown in (b,c,e). The dotted lines in (g,h) show the parabolic fits to the dark states emission energies (open circles).} \label{fig:Figure6}
\end{figure*}

\begin{equation}
\label{eqn:faraday_trion} E_{F}^{X^-,X^+}(B)=E_0 \pm \frac{\mu_{B}B}{2}\left | g_e^{sz}+g_h^{z}\right |+\gamma_{2}^{X^-,X^+}(0^{o})B^2.
\end{equation}

where $\gamma_{2}^{X^-}(\theta)$ and $\gamma_{2}^{X^+}(\theta)$ are the diamagnetic shifts of the negative and positive trions respectively. In Voigt geometry, the emission energies from the predominantly bright states are:\cite{Tischler2002}

\begin{equation}
\label{eqn:voigt_b_trion} E_{V,B}^{X^-,X^+}(B)=E_0 \pm \frac{\mu_{B}B}{2}\left | g_e^{sx}+g_h^{x}\right |+\gamma_{2}^{X^-,X^+}(90^{o})B^2,
\end{equation}

and those from the predominantly dark states are:

\begin{equation}
\label{eqn:voigt_d_trion} E_{V,D}^{X^{-},X^{+}}(B)=E_0 \pm  \frac{\mu_{B}B}{2}\left | g_e^{sx}-g_h^{x}\right |+\gamma_{2}^{X^-,X^+}(90^{o})B^2.
\end{equation}

The magnetic field dependence of $X^+$ and $X^-$ PL energies were fitted with the same approach than for $X^0$. It is worth emphasizing that experiments performed in Voigt geometry do not allow for an accurate determination of $g_e^x$ and $g_h^x$, the splittings between the different emission lines being too small. In view of this, experiments carried out at $\theta=45^o$ [Figure \ref{fig:Figure5}(b)]  yield a balance between the emission intensity of the predominantly dark states and their energy splitting (splitting of about 200 $\mu$eV at 8 T). From the magnetic-field dependence of $X^0$ and the trions for $\theta=0,15,30,45,60$ and $90^o$, we deduce $|g_e^{sx}|=0.40\pm0.01, |g_e^{sz}|=0.47\pm0.01, |g_h^{x}|= 0.02\pm0.01$ and $|g_h^{z}|=1.50\pm0.05$. First, we manage with this set of parameters to fit the $X^0$, $X^{+}$ and $X^{-}$ magneto-PL satisfactorily for all $\theta$. The value deduced for $|g_e|$ is typical of InAs quantum dots\cite{Krizhanovskii2005,Toft2007,Oberli2009,Sheng2008a} and the slight anisotropy obtained for the $g_e$ tensor confirms recent time-resolved pump-probe ellipticity experiments.\cite{Schwan2011} Although magneto-PL experiments can only provide the modulus of the electron $g$-factor (see discussion by Oberli {\itshape et al.} \cite{Oberli2009}), we assume that the $s$-electron $g$-factor is negative, in agreement with recent theoretical \cite{Pryor2006,Pryor2007} and experimental reports \cite{Maletinsky2007}. We therefore take $g_e^{sz}$ = -0.47 and $g_e^{sx}$ = -0.40. Concerning $g_h$, we attribute the small value of its in-plane component to the fact that the cubic terms in the spin Hamiltonian are negligible, as previously shown by Toft and Phillips.\cite{Toft2007}. We emphasize that the values reported in the literature for $g_h^z$ and $g_h^x$ in InAs QDs are extremely scattered.\cite{Krizhanovskii2005,Toft2007,Oberli2009,Schwan2011a} This is usually assigned to the fact that $g_h$ is a function of the heavy-hole light-hole mixing, the latter being strongly dependent on the shape, the composition and the strain state of the QD.\cite{Nakaoka2005,Babinski2006a,Sheng2007,Sheng2008b}

\subsection{Magneto-photoluminescence of doubly negatively charged excitons}

We present in this Section the emission properties of $X^{2-}$ in magnetic field for various $\theta$. We only discuss here the magnetic-dependence of the recombination of $X^{2-}$ into the triplet final state. The emission corresponding to the recombination into the singlet state is too weak for us to track its magnetic field-dependence. At zero magnetic field, the recombination into the triplet is composed of two lines (FWHM = $80\pm2$ $\mu$eV) split by 99 $\mu$eV (Figure \ref{fig:Figure4}). In contrast to the $X^0$, both the $|\pm1 \rangle$ and the $|\pm2 \rangle$ states of the $X^{2-}$ couple to the light at zero magnetic field (we still denote these states as bright and dark, respectively, in analogy with what done in Section III.B). The exchange interaction between the $s$-hole and the $p$-electron of the $X^{2-}$ should in principle make it possible to resolve four distinct lines for the recombination into the triplet: \emph{(i)} the isotropic exchange interaction splits energetically the bright and the dark states of $X^{2-}$ by an amount $c_z^{sp}/2$ and \emph{(ii)} the $|\pm1 \rangle$ states are further split by an amount that we take equal to $(c_x^{sp}+c_y^{sp})/2$ [see Figure \ref{fig:Figure6}(a)].  Following the approach by Kazimierczuk {\itshape et al.} and Ediger {\itshape et al.} \cite{Ediger2007b,Kazimierczuk2011}, we first neglect the exchange energy splitting between the $|\pm2 \rangle$ states. Then, and in contrast to the case of $X^0$, $(c_x^{sp}+c_y^{sp})/2$ is typically 1.4 to 1.7 times larger than $c_z^{sp}/2$.\cite{Ediger2007b,Kazimierczuk2011} We thus ascribe the lower energy line of the doublet to the recombination of the $|\pm2 \rangle$ and the $|-1 \rangle$ states, while the higher energy one corresponds to the emission from the $|+1 \rangle$ state.\cite{Ediger2007b} We also deduce $(c_x^{sp}+c_y^{sp})/2 = 99$ $\mu$eV and $c_z^{sp}/2=65\pm6$ $\mu$eV, in agreement with previous reports.\cite{Ediger2007b}

The $\theta$-dependent magneto-PL of the $X^{2-}$ is displayed in Figure \ref{fig:Figure6}(b-e). In Faraday geometry, applying a magnetic field splits both the initial and the final states involved in the $X^{2-}$ recombination [Figure \ref{fig:Figure6}(a)]. As a first approximation, following Kazimierczuk {\itshape et al.},\cite{Kazimierczuk2011} we consider similar $g$-factors for electrons in the $s$ and the $p$-shells, irrespectively of the presence of a hole in the $s$-shell. It follows that the energy of the photons resulting from the recombination of the $| -1 \rangle$ and the $| +2 \rangle$ states [labeled as transitions 1 and 2 in Figure 6(a)] should exhibit a Zeeman shift of $\mu_B|g_h^z-g_e^{sz}|B/2$. Similarly, we obtain that the linear shift of the emission from the $| -2 \rangle$ and the $| +1 \rangle$ $X^{2-}$[transitions 4 and 3 in Figure 6(a)] would be $-\mu_B|g_e^{sz}-g_h^z|B/2$. Since the exchange terms involved are smaller than the spectral resolution of our set-up, the emission from the  $| +2 \rangle$ and $| -1 \rangle$ ($| -2 \rangle$ and $| +1 \rangle$, respectively) $X^{2-}$ overlap: we therefore only observe two distinct lines in Faraday geometry [Figure 6 (b)]. Now, the situation is made complicated by the fact that the $p$-electron envelope spreads into the GaAs barriers, this spreading depending on the presence of the hole in the $s$-shell (see Section III.A). Consequently, not only should the $s$- and the $p$-electron $g$-factors differ,\cite{Akimov2005,Alegre2006,Kazimierczuk2011b} but so will the $p$-electron $g$-factor for the initial and the final states of the $X^{2-}$ recombination.\cite{Kiselev1998} With $g_e^{pz,i}$ and $g_e^{pz,f}$ the $p$-electron $g$-factors of the initial and the final states of the $X^{2-}$ recombination, the magnetic field dependence of the recombination energies of the $X^{2-}$ dark states in Faraday geometry are given by:

\begin{equation}
\label{eqn:faraday_X2a}
\begin{split}
E_{F,D}^{X^{2-}}(B)&=E_0 - \frac{c_z^{sp}}{4}  \pm \frac{\mu_{B}B}{2}|g_e^{sz}-g_h^{z}+g_e^{pz,f}-g_e^{pz,i}|  \\
&+\gamma_{2}^{X^{2-}}(0^{o})B^2
\end{split}
\end{equation}

where $g_e^{pz,i}$ and $g_e^{pz,f}$ are the $p$-electron $g$-factors for the initial and the final states of the $X^{2-}$ recombination. In a similar manner, the emission energies for the recombination of the bright states into the triplet are:

\begin{equation}
\label{eqn:faraday_X2b}
\begin{split}
E_{F,B}^{X^{2-}}(B)&=E_0 + \frac{c_z^{sp}}{4} \pm \frac{1}{2}\mu_BB|g_e^{pz,f}-g_e^{sz}|+\gamma_{2}^{X^{2-}}(0^{o})B^2\\ &\pm \frac{1}{2}\sqrt{\frac{[\mp(c_x^{sp}+c_y^{sp})]^2}{4}+(\mu_{B}B|g_h^{z}-g_e^{pz,i}|)^2}
\end{split}
\end{equation}

Using Equations (\ref{eqn:faraday_X2a}) and (\ref{eqn:faraday_X2b}) to fit the data in Figure 6(a) (we take $g_e^{sz}$, $g_h^z$, $c_z^{sp}$, $(c_x^{sp}+c_x^{sp})/2$ and $\gamma_2^{X^{2-}}(\theta)$ as determined above), we get  $|g_e^{pz,f}-g_e^{pz,i}| = 0.12$ [the result of the fitting is shown in Figure 6(f)]. In agreement with our discussion in Section IV.A on the diamagnetic coefficient of the $X^{2-}$ complex, we attribute the observed variation in $p$-electron $g$-factor to the increased spreading of the $p$-electron wavefunction inside the dot barriers after recombination of the $X^{2-}$. As the $p$-electron is more bound to the dot for the initial state of the $X^{2-}$, its $g$-factor should be close to that measured for the $s$-electron. As an approximation, we first take $g_e^{pz,i}=g_e^{sz}$ = -0.47, which leads to $g_e^{pz,f}$ = -0.59 or $g_e^{pz,f}$ = -0.35. Second, as the $p$-electron wavefunction probes both the QD and the barrier, $g_e^{pz,f}$ ranges between the electron $g$-factor in the QD ($g_e^{sz}$ = -0.47) and the electron $g$-factor in the barrier. In the presence of alloy intermixing between the InAs QD and GaAs barriers,\cite{Vastola2012} the latter is between $g_e^{InAs}$ = -14.7 and $g_e^{GaAs}$ = -0.44,\cite{Konopka1967} which leads finally to $g_e^{pz,f}$ = -0.59.

When a magnetic field with an in-plane component is applied, the initial and the final states of the $X^{2-}$ recombination get mixed. Accounting for the different channels available for the recombination of each of the four $X^{2-}$ states, we could now expect to detect twelve distinct emission lines. Now, as mentioned above, the exchange terms are small compared to our spectral resolution. Furthermore, not all the possible transitions carry an oscillator strength large enough for detection.\cite{Kazimierczuk2011} We therefore only detect four distinct lines [Figure \ref{fig:Figure6}(b,c)], although, based on the analysis in Ref. \onlinecite{Kazimierczuk2011}, we suspect each of these lines to be the result of the convolution between the emission from two distinct $X^{2-}$ recombination channels. Similarly to what seen for $X^0$ and the trions and in agreement with magneto-PL experiments on $X^{2-}$ in CdTe/ZnTe QDs,\cite{Kazimierczuk2011} the Zeeman splittings reduce with increasing $\theta$. In Voigt geometry, it becomes almost impossible to resolve the multiple splittings of the emission lines [Figure \ref{fig:Figure6}(d)]. In addition, any quantitative discussion of the Zeeman splittings at intermediate $\theta$ is hindered by the fact that (\emph{i}) $g_e^{pz}$ and $g_e^{px}$ depends on the spreading of the $p$-electron wave function in the barriers and (\emph{ii}) the shape of the $p$-electron wave function strongly depends on the field orientation.\cite{Alegre2006}

\section{Angle-dependent hybridization of electronic quantum dot states with the wetting layer}

\begin{figure*}[h!t]
  \includegraphics[width=\textwidth]{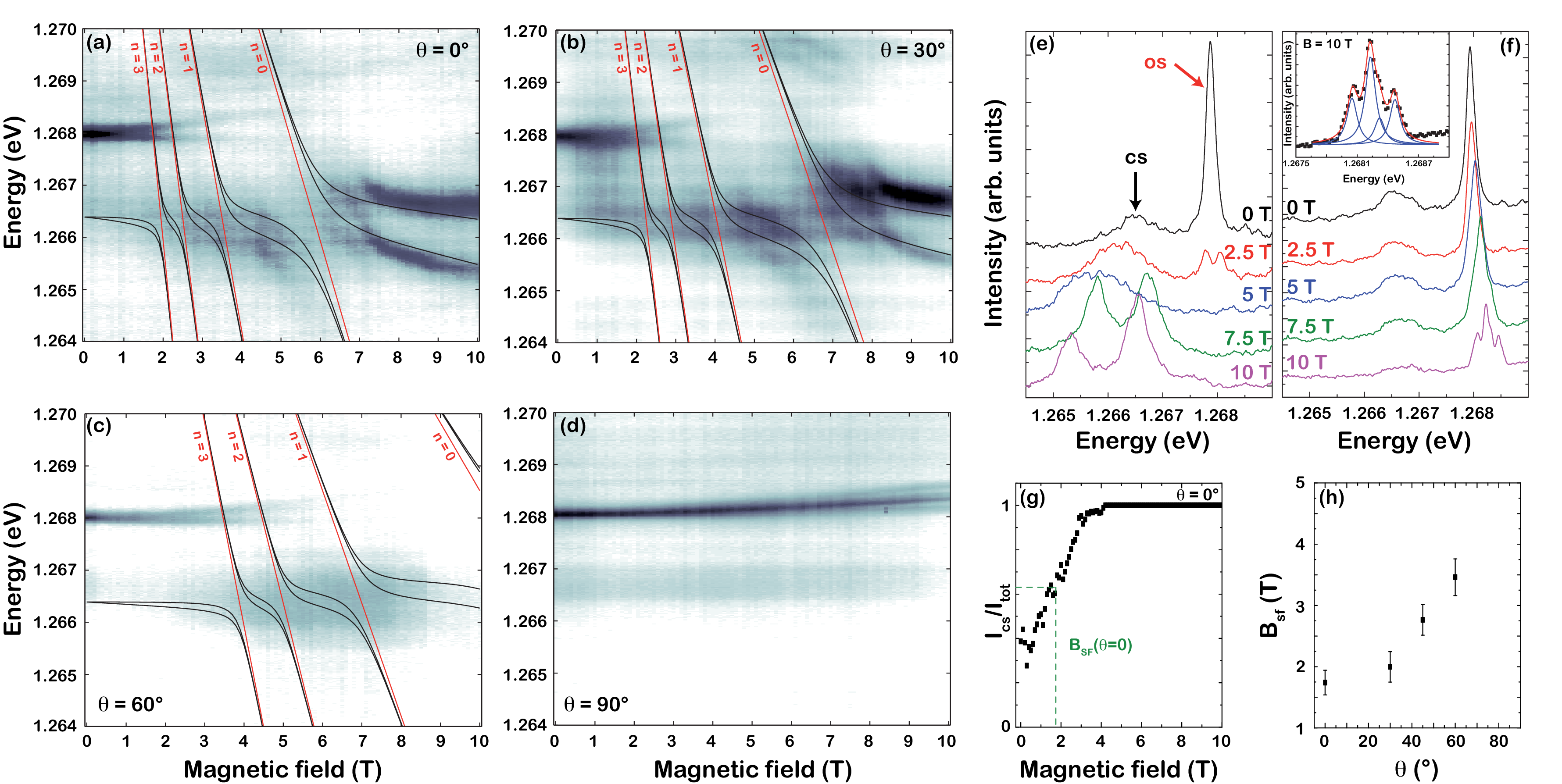}
  \caption{(colour online): (a)-(d) Magneto-photoluminescence intensity maps versus photon energy and magnetic field of the $X^{3-}$ taken at $\theta= $ $0^o$ (a), $30^o$ (b), $60^o$ (c) and $90^o$ (d). Solid lines in (a-c) are guides to the eye highlighting the anticrossings that result from the coupling between the final state of the $X^{3-}$ recombination and the electron Landau levels of the wetting layer. (e,f) PL spectra of the $X^{3-}$ taken at 0, 2.5, 5, 7.5 and 10 T in Faraday (e) and Voigt (f) geometries. In (e) the black and red arrows indicate the emission from the closed (cs) and open-shell (os) $X^{3-}$. The inset in (f) shows that the emission of the $X^{3-}$ in Voigt geometry splits into four components. The red line shows the result of a deconvolution of the spectrum using four Lorentzian curves with a FWHM of 120 $\mu$eV (blues lines). (g) Ratio between the emission intensity from the closed-shell $X^{3-}$ $I_{(cs)} $ and the total $X^{3-}$ $I_{(tot)} $ emission intensity with respect to the magnetic field. We define as $B_{sf}$ the magnetic field at which $\frac{I_{cs}}{I_{tot}}=1-(1/e)$ (green dashed line). (h) Evolution of $B_{sf}$ with respect to $\theta$.}
\label{fig:Figure7}
\end{figure*}

We now turn our attention to the magneto-PL from the triply negatively charged exciton $X^{3-}$. The two $p$-electrons of this complex may occupy either a different or the same sub-shell. In the case of QDs with in-plane isotropy, the $p_x$ and $p_y$ states are degenerate and the $X^{3-}$ exhibits an open-shell configuration: both the $p_x$ and the $p_y$ states are occupied and the two $p$-electrons have parallel spins. The open-shell $X^{3-}$ may recombine into a triplet or a singlet state, giving rise to two emission bands split by electron-electron exchange interaction.\cite{Warburton2000} As was the case for the $X^{2-}$, we will not comment further on the recombination of the $X^{3-}$ into the singlet because of the relatively low intensity of this emission.\cite{Warburton2000} In the presence of in-plane anisotropy, the degeneracy between the $p_x$ and the $p_y$ states is lifted. When the in-plane anisotropy is such that the energy separation between the $p_x$ and the $p_y$ states is larger than the exchange energy between the $p_x$ and $p_y$ electrons, both $p$-electrons occupy the lower-energy subshell:\cite{Urbaszek2003} the $X^{3-}$ is said to adopt a closed-shell configuration. Note that the observation at zero magnetic field of the closed-shell $X^{3-}$ does not necessarily imply that the QD exhibits an in-plane anisotropy. For instance, the presence of electrons in the wetting layer is sufficient to break the potential symmetry in the QD and to favour the closed-shell configuration.\cite{Warburton2003,Warburton2005}

\begin{figure}[h]
\includegraphics[scale=0.35]{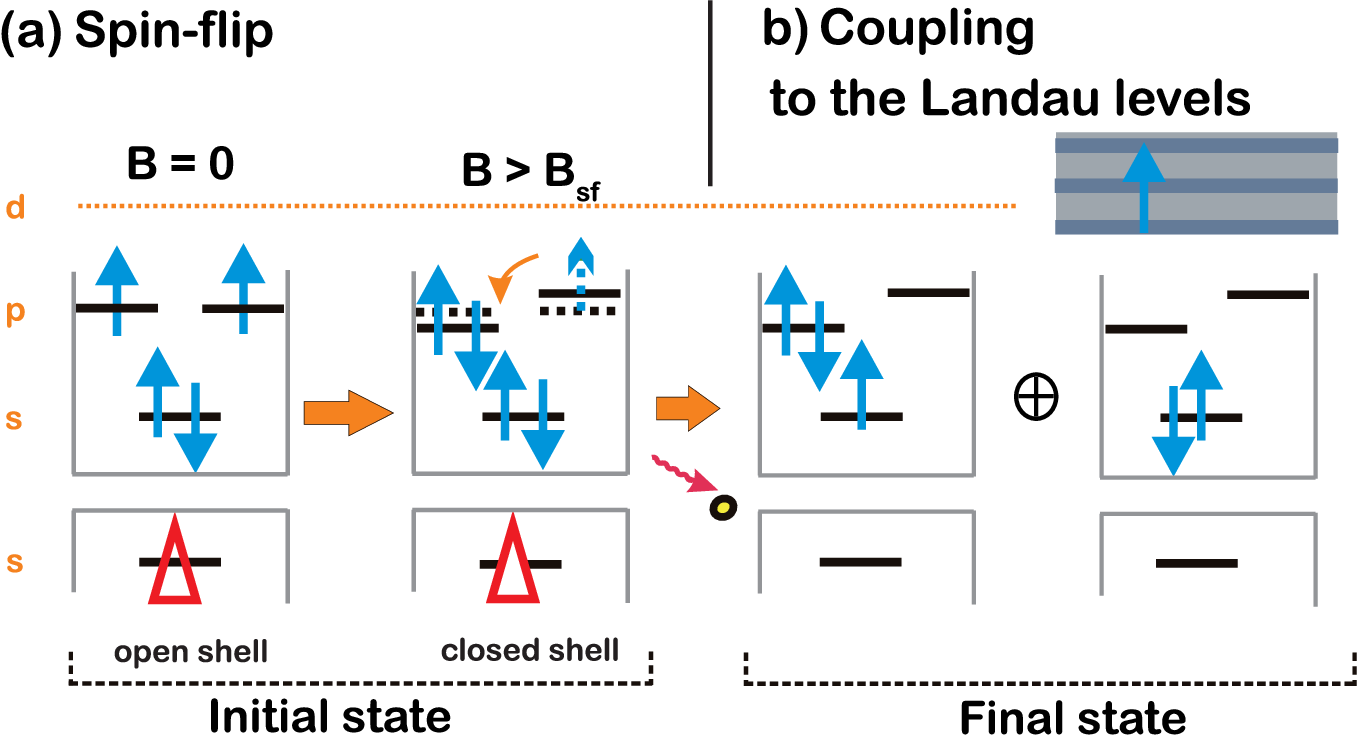}
\caption {(colour online) (a) Applying a magnetic field parallel to the quantum dot high symmetry axis splits the $p_x$ and $p_y$ subshells, favouring the closed-shell $X^{3-}$ against the open-shell one. (b) After recombination of the closed-shell $X^{3-}$, the QD is in an excited state. Auger interaction promotes an electron into the $d$-shell, which couples with wetting layer states. In presence of magnetic field, the wetting layer continuum of states develops into Landau levels, giving rise to the anticrossings seen in Figure \ref{fig:Figure7}(a-c). This figure follows the scheme outlined by Karrai \emph{et al.}\cite{Karrai2004}} \label{fig:Figure8}
\end{figure}

The emission spectra from the $X^{3-}$ at zero-magnetic field is shown in Figure \ref{fig:Figure7}. It is dominated by an emission band centered at 1267.9 meV that we attribute to the recombination of the open-shell $X^{3-}$ into the triplet. The open-shell $X^{3-}$ emission shows a fine structure (Figure \ref{fig:Figure4}), as a result of exchange interaction between the $s$-hole and the $p$-electrons.\cite{Urbaszek2003,Warburton2005} While the 100 $\mu$eV exchange splitting observed here between the open-shell $X^{3-}$ states is similar to what reported respectively by Urbaszek {\itshape et al.} and Warburton {\itshape et al.} \cite{Urbaszek2003,Warburton2005} (Figure \ref{fig:Figure4}), the emission lines appeared to be much broader (FWHM = 170 $\mu$eV compared with 70 $\mu$eV in Refs. \onlinecite{Urbaszek2003,Warburton2005}). We observe a much weaker and broader (FWHM = 790 $\mu$eV) emission band centered at 1266.5 meV [Figure \ref{fig:Figure7}(e)], that we ascribe to the recombination of the closed-shell $X^{3-}$.\cite{Govorov2004} As shown in Figure \ref{fig:Figure8}, after recombination of the closed-shell $X^{3-}$, the QD is in an excited state. Fast reconfiguration of the final state occurs by an Auger process and results in the large emission broadening observed in Figure \ref{fig:Figure7}. We emphasise that the electron excited into the $d$-shell couples to the wetting layer.\cite{Govorov2003,Govorov2004} The behaviour of the magneto-PL of the closed-shell $X^{3-}$, the so-called Kondo-Anderson excitons,\cite{Govorov2003,Govorov2004b} will be discussed in the remainder of this paper.

The evolution of the $X^{3-}$ emission energies in a magnetic field at $\theta=0$ is quite different from that observed so far for less-charged complexes [Figure \ref{fig:Figure7}(a)]. First, when increasing the magnetic field, the emission from the open-shell $X^{3-}$ quenches in favour of the closed-shell emission. Applying a magnetic field splits the $p_x$ and $p_y$ subshells, making the closed-shell configuration energetically more favourable.\cite{Warburton2003} With $I_{cs}$ and $I_{tot}$ the emission intensity from the closed-shell $X^{3-}$ and the total $X^{3-}$ emission intensity, respectively, we define the spin-flip magnetic field $B_{sf}$ as the field at which $\frac{I_{cs}}{I_{tot}}=1-\frac{1}{e}\simeq 0.63$.\footnote{This definition for $B_{sf}$ derives from the assumption that $\frac{I_{cs}}{I_{os}}\propto \exp(\Delta E/k_BT)$, where ${I_{os}}$ is the intensity of the emission from the open-shell $X^{3-}$, $k_B$ is the Boltzmann constant, and where the energy difference $\Delta E\propto B\cos\theta$ between open and closed-shell configurations scales linearly with B} We measure $B_{sf}= 1.77\pm 0.1$ T [Figure \ref{fig:Figure7}(g)]]. Second, the emission from the closed-shell $X^{3-}$ does not follow a parabolic dependence with increasing magnetic field [Figure \ref{fig:Figure7}(a,e)]. In agreement with earlier works,\cite{Warburton2003,Karrai2004,Warburton2005} the closed-shell $X^{3-}$ emission shows several anticrossings for intermediate values of the magnetic field ranging between 2 and 8 T. This is the signature of the hybridization between localized (QD with a full $p$-shell and one electron in the $s$-shell) and extended states (QD with a full $s$-shell and one electron delocalized in the wetting layer).\cite{Karrai2004} We underline that the asymptotes of the closed-shell $X^{3-}$ emission energies are linear with the magnetic field, as a result of the development of the wetting layer continuum of states into Landau levels. At resonance between the localized and the delocalized electronic states, $\Delta =(n+3/2)e\hbar B/m_e$, where $m_e$ is the electron effective mass in the wetting layer, $n$ the index for the $n^{th}$ Landau level and $\Delta$ the kinetic energy at zero field of the electron in the wetting layer.\cite{Warburton2003} Using $m_e=0.067m_0$, we obtain $\Delta =15\pm1$ $\mu$eV. Concerning the asymptotes themselves, their magnetic field dependences are given by:\cite{Karrai2004}

\begin{equation}
\label{eqn:asymptote} E_0+\Delta - (n+\frac{3}{2})e\hbar B / m_e,
\end{equation}

where $E_0$ is the emission energy from the closed-shell $X^{3-}$ at zero field. For magnetic fields larger than 8 T, the final state of the $X^{3-}$ recombination is fully localized in the QD. As the initial and the final states of the recombination are influenced by the magnetic field only through coupling to the spins of the $s$-hole and the $s$-electron, respectively, the closed-shell emission is split into two components whose energy separation is proportional to $|g_e^{sz}-g_h^z|B$. In addition, we deduce from Figure \ref{fig:Figure7}(e) a paramagnetic behaviour (i.e. $\gamma_2<0$) for the closed-shell $X^{3-}$. According to our previous discussions in Sections IV.A and IV.C, this observation confirms that the extent into the GaAs barriers of the $p$-electron envelope is larger for the final state than for the initial state of the recombination.

Some remarkable changes occur when the magnetic field is tilted with respect to the growth axis. First, as displayed in Figure \ref{fig:Figure7}(a-d,h), $B_{sf}$ increases with $\theta$. We emphasize that in Voigt geometry, the emission from the open-shell $X^{3-}$ dominates the spectrum over the whole range of investigated magnetic fields. To explain this observation, we consider the evolution of the energy spectrum of a QD in a tilted magnetic field. This generalised version of the Fock-Darwin problem has been solved analytically by  Henriques and Efros \cite{Henriques2009}. In this work, Henriques and Efros predict for a given value of the magnetic field a reduced splitting between the $p_x$ and $p_y$ states as $\theta$ is increased. It follows that $B_{sf}$ increases with $\theta$, in agreement with the data in Figure \ref{fig:Figure7}(h). Second, the slope of the asymptotes of the closed-shell $X^{3-}$ emission energies decrease with $\theta$ [Figure \ref{fig:Figure7}(a-c)]. In particular, when $\theta=90^o$, the magnetic field-dependence of the closed-shell $X^{3-}$ emission energies exhibits no anticrossing [Figure \ref{fig:Figure7}(d,f)]. We attribute such a decrease in the slopes of the asymptotes to the fact that the eigenenergies for an electron in a quantum well in presence of a tilted magnetic field are proportional to the component of the magnetic field parallel to the confinement axis.\cite{Bastard1988} In agreement, the asymptotes in Figure \ref{fig:Figure7}(b,c) are fitted fairly well using Equation (9) with $B$ the on-axis component of the applied magnetic field. We underline that tuning  $\theta$ and the amplitude of the magnetic field  modifies the shape of the wave function of electrons confined in the QD,\cite{Alegre2006} which opens the possibility of studying different configurations of Kondo-Anderson states. When $\theta=90^o$, both the open-shell and the closed-shell $X^{3-}$ emission energies exhibit a quadratic behaviour with magnetic field, as an indication that the $X^{3-}$ can now be treated in the artificial atom model. For instance, at 10 T, the open-shell $X^{3-}$ emission splits into four lines, as a result of the Zeeman effect on both the initial and the final state of the recombination. Concerning, the $X^{3-}$ in the closed-shell configuration, its emission energy in Voigt geometry shows almost no dependence on the magnetic field: no splitting is resolved and $\gamma_2 = 0.5$ $\mu$eV/$T^2$.

\section{Conclusions}

In conclusion, we have carried out photoluminescence experiments in tilted magnetic field on neutral and charged excitons in InAs/GaAs quantum dots. Performing magneto-photoluminescence with different field orientation not only allows estimation of the shape of the exciton wave function, but also makes it possible to determine accurately the Land\'e $g$-tensors for electrons and holes in the $s$-shell. Experiments on doubly and triply negatively charged excitons provide information on carriers in high-energy shells.  In particular, we have demonstrated that the $g$-factor of the $p$-electron depends strongly on the spreading of its wave function inside the dot barriers and on the presence of the hole in the $s$-shell. Finally, we have observed coherent coupling between the quantum dot electronic states and the wetting layer Landau levels. We have investigated the magnetic field-dependence of this coupling for various orientations of the magnetic field. Based on the energy spectrum of electrons in quantum dot in tilted magnetic fields, we have shown that when increasing the angle between the magnetic field and the dot high symmetry axis, the system evolves from a Kondo-Anderson exciton description to the artificial atom model.

\begin{acknowledgments}

We acknowledge funding from the EPSRC. B.V.H also thanks the Hitachi Cambridge Laboratory for additional funding. P.C. acknowledges financial support from the European Union Seventh Framework Programme under Grant Agreement No. 265073.
\end{acknowledgments}

\bibliography{bibliography}

\end{document}